\documentclass[11pt,a4paper]{article}
\usepackage{amsmath,amsthm,amssymb,epsfig,latexsym}
\usepackage{epsfig}
\usepackage[T1]{fontenc}    
\usepackage{graphics}
\usepackage{graphicx}
\usepackage{pstricks,pst-coil,pst-fill,pst-plot}
\newcommand{\be}[1]{\begin{equation}\label{#1}}
\newcommand{\ba}[1]{\begin{multline}\label{#1}}
\newcommand{\ee}{\end{equation}}
\newcommand{\ea}{\end{eqnarray}}

\newcommand{\num}{\\\rule{0pt}{20pt}}
\newcommand{\numa}[1]{\\\rule{0pt}{#1pt}}

\newcommand{\tr}{\mathop{\rm tr}}

\newcommand{\qp}{s^{\scriptscriptstyle +}}
\newcommand{\qm}{s^{\scriptscriptstyle -}}
\newcommand{\qpm}{s^{\scriptscriptstyle \pm}}
\newcommand{\qmp}{s^{\scriptscriptstyle \mp}}
\newcommand{\Rp}{r^{\scriptscriptstyle +}}
\newcommand{\Rm}{r^{\scriptscriptstyle -}}
\newcommand{\Rpm}{r^{\scriptscriptstyle \pm}}

\newcommand{\np}{n_{\scriptscriptstyle +}}
\newcommand{\npm}{n_{\scriptscriptstyle \pm}}
\newcommand{\nmp}{n_{\scriptscriptstyle \mp}}
\newcommand{\Dp}{D^{\scriptscriptstyle +}}
\newcommand{\Cp}{C^{\scriptscriptstyle +}}
\newcommand{\Dpm}{D^{\scriptscriptstyle \pm}}
\newcommand{\Cpm}{C^{\scriptscriptstyle \pm}}
\newcommand{\nm}{n_{\scriptscriptstyle -}}
\newcommand{\Dm}{D^{\scriptscriptstyle -}}
\newcommand{\Cm}{C^{\scriptscriptstyle -}}

\newtheorem{thm}{Theorem}[section]

\newtheorem{lemma}{Lemma}[section]
\newtheorem{cor}{Corollary}[section]

\def\qed{\hfill\nobreak\hbox{$\square$}\par\medbreak}
 \makeatletter
 \@addtoreset{equation}{section}
 \makeatother
 
\begin{document}

\vspace{14pt}

\begin{center}
\begin{LARGE}
\vspace*{1cm} {Integral operators with the generalized sine-kernel\\
on the real axis}
\end{LARGE}

\vspace{20pt}

\begin{large}

{\bf N.~A.~Slavnov}\footnote[1]{ Steklov Mathematical Institute,
Moscow, Russia, nslavnov@mi.ras.ru}

\end{large}

\vspace{30pt}

\vspace{1cm}
\parbox{12cm}{\small The asymptotic properties of integral
operators with the generalized sine kernel acting on the real axis
are studied. The formulas for the resolvent and the Fredholm
determinant are obtained in the large $x$ limit. Some applications
of the results obtained to the theory of integrable models are
considered.}

\vspace{20pt}

\centerline{Keywords:  Fredholm determinant, resolvent, asymptotic
expansion}

\end{center}

\vspace{30pt}

\section{Introduction\label{INTR}}

Integral operators with a sine kernel arise in many areas of
mathematical physics. The sine kernel has the form
\begin{equation}\label{ISK}
 S(\lambda,\mu)= \frac{\sin \frac{x}2(\lambda-\mu)}{\pi(\lambda-\mu)} ,
\end{equation}
where $\lambda$ and $\mu$ are integration variables, $x$ is a real
parameter. The operator $I+\gamma S$, where $I$ is the identity
operator and $\gamma$ is a complex number, acts on an interval (or a
system of intervals) $\ell$ of the real axis on functions from
$L_2(\ell)$.

The Fredholm determinant of the integral operator $I-S$  appears in
random matrix theory \cite{Gau61}. In the scaling limit,
${\det}_{\ell}(I-S)$  gives the probability that a matrix belonging
to the Gaussian unitary ensemble has no eigenvalues in the interval
$x\ell$ \cite{GauM61}. The kernel \eqref{ISK} also appears in the
theory of quantum integrable systems. In particular, the determinant
${\det}_{\ell}(I+\gamma S)$ describes various zero temperature
correlation functions  of the impenetrable Bose gas model
\cite{Len64,JimMMS80}.

In the interpretations  of the sine kernel mentioned above the most
interesting question is related to the large $x$ behavior of the
Fredholm determinant $\det_\ell(I+\gamma S)$. This problem was
studied in numerous works (see e.g. \cite{Gau61,GauM61},
\cite{DesClo73}--\cite{DeiIKZ07}). A very nice connection of the
sine kernel to the Painlev\'e V equation was  investigated in
\cite{JimMMS80,DeiIZ97}. The methods of the asymptotic analysis of
$\det_\ell(I+\gamma S)$ are based on the fact that at $x$ large the
kernel $S$ becomes rapidly oscillating. Note however that the kernel
has no saddle points, and shifts of the integration contour do not
make it exponentially small. Therefore standard methods of
asymptotic estimates of oscillating integrals fail in this case.

Various generalizations of the  kernel \eqref{ISK} preserving its
oscillating structure are known in the literature. One of such
generalizations arises in the theory of truncated Wiener--Hopf
operators \cite{Foc44,NobL58}, where the integral operator acts on
$\mathbb{R}$ and the complex number $\gamma$ is replaced by some
function $\gamma(\lambda)$. Other generalizations of the sine kernel
were basically  used for the description of correlation functions of
matrix models or quantum integrable systems equivalent to free
fermions (see e.g. \cite{McCPS83}--\cite{CheZ04}). Most of these
kernels can be presented in the form
 \begin{equation}
 V(\lambda,\mu)=\frac{\sqrt{F(\lambda)F(\mu)} }{2i\pi (\lambda-\mu)}
  \bigl[ e_+(\lambda)e_-(\mu)-e_-(\lambda)e_+(\mu) \bigr], \quad
 \label{IGSK}
\end{equation}
where
 \begin{equation}\label{Iepm}
 e_\pm(\lambda)=\exp\left(\pm\frac{ixp(\lambda)}2\pm\frac{g(\lambda)}2\right),
 \end{equation}
and the operator $I+\gamma V$ acts on some contour ${\cal C}$. We
call the kernel \eqref{IGSK} the generalized sine kernel (GSK). The
large $x$ asymptotic behavior of the Fredholm determinant
${\det}_{\cal C}(I+\gamma V)$ where studied in the works mentioned
above for certain particular choices of functions $F$, $p$, and $g$
and the contour ${\cal C}$. It was shown in \cite{KitKMST09b} that
the Fredholm determinant asymptotic expansion  depends only on the
analytic properties of the functions $F$, $p$, and $g$ in an
neighborhood of the contour ${\cal C}$. Knowing these properties one
can evaluate the asymptotic behavior without using the explicit form
of these functions.

In the present article we focus our attention on a particular case
of the GSK where the integral operator $I+\gamma V$ acts on the
whole real axis. We also set $p(\lambda)=\lambda$, because just this
situation occurs in the most of applications. This case is more
simple than the one of a finite interval (moreover a union of finite
intervals). The matter is that for a finite interval $\ell$ the
large $x$ asymptotic behavior of ${\det}_\ell(I+\gamma V)$ mostly
follows from the analytic properties of the resolvent to the
operator $I+\gamma V$ in vicinities of the endpoints of $\ell$.
Although the methods of such analysis are now well developed (see
e.g. \cite{DeiIKZ07,CheZ04}), they are rather complicated
technically. As a result even in the simplest cases usually it is
possible to obtain explicitly only first several terms  of the
asymptotic expansion.

In the case of operators acting on $\mathbb{R}$  this difficulty
disappears due to the absence of the endpoints. Due to this one can
obtain uniform asymptotic estimates for the resolvent on the whole
real axis up to exponentially small corrections. As a result it is
possible to promote much further in studying of the Fredholm
determinant asymptotic behavior. In some particular cases  one can
obtain even the complete asymptotic expansion.

Similar picture arises in the case of the GSK acting on closed
contours $\cal C$. For these kernels the function $p(\lambda)$
usually is such that $|e^{ixp(\lambda)}|=1$ for $\lambda\in{\cal
C}$. The methods of the asymptotic analysis proposed in this paper
can be used in such cases as well.

This article is organized as follows. In
section~\ref{sec-pb-results}, we  announce the main results of the
paper, namely, the asymptotic formula for the resolvent of the
operator $I+\gamma V$ and two types of the asymptotic expansion of
${\det}_{\mathbb{R}}(I+\gamma V)$. The proofs of these results are
given in sections~\ref{Asy-form-Res}--\ref{S-Main-thm-FD1}. In
section~\ref{Ex} we consider two applications of the results
obtained.

\section{Notations and main theorems \label{sec-pb-results}}

In this section we specify some properties of the functions
$g(\lambda)$ and $F(\lambda)$ entering the kernel \eqref{IGSK}.
Recall that we have set $p(\lambda)=\lambda$.

\begin{figure}[h]
\begin{center}
 \begin{pspicture}(15,5)
 \psline[linestyle=dashed, dash=10pt 2pt]{->}(0.2,2)(14.4,2)
 \pscurve[linestyle=dashed, dash=1pt 1pt](1,2)(2,2)(3,1.9)(3.5,1.9)(3.7,1.9)(4.2,1.9)(4.3,1.9)(4.5,1.85)
 (4.7,1.5)(4.9,0.5)(5.1,0.55)(5.35,1.5)(5.5,2.9)(5.75,4.05)(6.1,4.9)
 (6.2,5)(6.55,4.5)(6.8,3.2)(7,2.2)(7.1,2.1)(7.3,2.05)(8,2.05)(12,2.05)
 %
 \pscurve(1,2)(2,2)(3,2)(3.5,2.5)(3.7,4)(4.2,2.5)(4.3,2)(4.4,1.85)(4.6,1.5)(4.8,0.5)(5,0.5)(5.2,1.5)
 (5.3,1.9)(5.5,2)(6.4,2.3)(6.6,3.8)(6.7,3.95)(7.1,3.9)(7.2,3.1)(7.4,2.1)(7.5,2)
 (7.9,2)(9,2)(9.4,1.85)(9.6,1.5)(9.7,0.57)(9.75,0.58)(9.85,0.65)(9.9,0.75)(10.2,1.5)(10.3,1.9)(10.6,2)(11,2)(12,2)
 \rput(4.5,4.2){$\Rp_1$}%
 \rput(4.1,3.9){$\scriptstyle\circ$}
 \rput(4.5,3.5){$\qp_1$}%
 \psdots(3.9,3.5)
 \rput(5.5,1){$\qm_1$}
 \rput(4.3,1){$\Rm_1$}
 \rput(4.3,0.7){$\scriptstyle\circ$}%
 \psdots(5,1)
 \rput(4.8,2.6){$\Gamma^{(2)}_{13;12}$}
 \rput(5.3,3.9){$\Gamma^{(1)}_{2;1}$}
 \rput(14.2,2.2){$\mathbb{R}$}
 \rput(7.4,4.3){$\Rp_3$}%
 \rput(7.05,4.2){$\scriptstyle\circ$}
 \rput(5.65,5.1){$\qp_2$}%
 \psdots(6.3,4.8)
 \rput(6.9,3.5){$\qp_3$}
 \rput(6.8,4.99){$\Rp_2$}
 \rput(6.55,4.8){$\scriptstyle\circ$}%
 \psdots(6.9,3.8)
 \rput(8.8,1.2){$\Rm_2$}%
 \rput(8.5,1.1){$\scriptstyle\circ$}
 \rput(9.6,0.35){$\qm_2$}%
 \psdots(9.8,0.8)
 \rput(10.5,1.05){$\qm_3$}
 \rput(11.8,1){$\Rm_3$}
 \rput(11.42,0.8){$\scriptstyle\circ$}%
 \psdots(10.62,0.65)
 %
 \psline[linewidth=2pt]{<-}(4.52,1.65)(4.49,1.7)
 \psline[linewidth=2pt]{<-}(5.69,3.7)(5.67,3.65)
%
\end{pspicture}
\caption{\label{G1212}{\small The poles  $\Rpm_j$ are shown by
$\scriptstyle\circ$, the zeros $\qpm_j$ of $1+\gamma F(\lambda)=0$
are shown by $\scriptstyle\bullet$. The inte\-gra\-ti\-on contour
$\Gamma^{(2)}_{13;12}$ (solid line) bypasses the points $\qp_1$,
$\qp_3$ from above and $\qm_1$, $\qm_2$ from below. The contour
$\Gamma^{(1)}_{2;1}$ (dotted line) bypasses the point $\qp_2$  from
above and $\qm_1$ from below. }}
\end{center}
\end{figure}
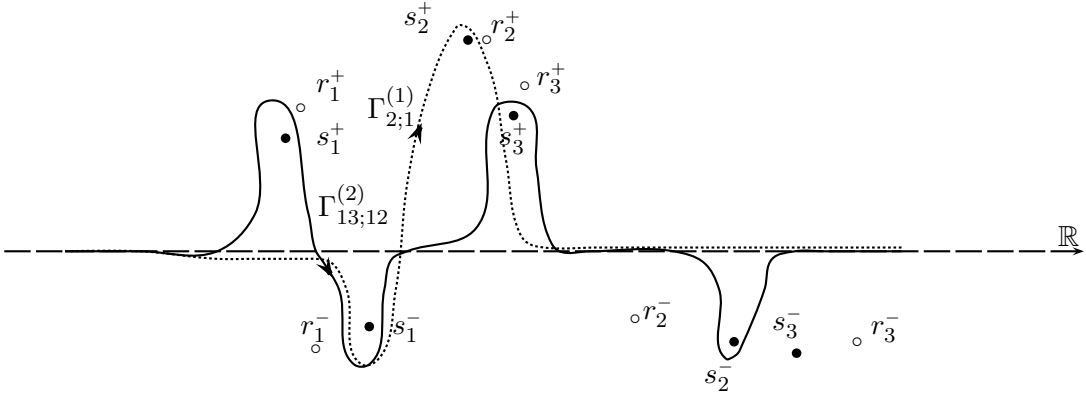

Let $g(\lambda)$ be holomorphic in a strip $|\Im(\lambda)|<a$,
$a>0$. Assume also that $g(\lambda)/\lambda\to 0$ as
$\Re(\lambda)\to\pm\infty$ and $|\Im(\lambda)|<a$. The function
$F(\lambda)$ is holomorphic in the same strip except $\np+\nm$
simple poles in the points $\Rpm_j$ separated from
the real axis\footnote[1]{%
The case of higher order poles can be analyzed by taking the limits
$\Rpm_j\to\Rpm_k$. We do not consider this case, because the
explicit formulas become much more cumbersome.}. We assume that the
points $\Rp_j$, $j=1,\dots,\np$ lie in the upper half-plane and
$\Rm_k$, $k=1,\dots,\nm$ lie in the lower half-plane (see
Fig.~\ref{G1212}). The complex number $\gamma$ is supposed to be
small enough, such that  zeros $\qpm_j$ of the function $1+\gamma
F(\lambda)$ are slightly shifted from the poles $\Rpm_j$ (see
Fig.~\ref{G1212}). The zeros  $\qp_j$, $j=1,\dots,\np$ belong to the
upper half-plane and $\qm_k$, $k=1,\dots,\nm$ belong to the lower
half-plane. Finally we assume that $F(\lambda)\to 0$ as
$\Re(\lambda)\to\pm\infty$, $|\Im(\lambda)|<a$ in such a way that
$|\tr V|<\infty$.

%
%
Under the conditions listed above  the Fredholm determinant
${\det}_{\mathbb{R}}(I+\gamma V)$  exists and it is an entire
function of $\gamma$. Since at $\gamma=0$ this determinant is equal
to $1$, we conclude that ${\det}_{\mathbb{R}}(I+\gamma V)\ne 0$, if
$\gamma$ belongs to some vicinity of the origin. Hence, the
resolvent to the operator $I+\gamma V$ exists at least for $\gamma$
small enough.

Let us introduce now several functions used below. First, we define
functions $\nu(\lambda)$ and $\alpha(\lambda)$ as
 \begin{equation}\label{nu-1}
 \nu(\lambda)=\frac{-1}{2\pi i}\log \bigl(1+\gamma F(\lambda)\bigr), \qquad
 \alpha(\lambda)\equiv
 \alpha([\nu],\lambda)=\exp\biggl(\;\int\limits_{\mathbb{R}}
 \frac{\nu(\mu)\,d\mu}{\mu-\lambda}\biggr).
 \end{equation}
The branch of the logarithm in \eqref{nu-1} is fixed by the
condition $\nu(+\infty)=0$. Clearly $\alpha(\lambda)$ has a cut on
the real axis, and its limiting values $\alpha_\pm(\lambda)$ from
the upper and lower half-planes enjoy the property
 \begin{equation}\label{RHP-a}
 \alpha_-(\lambda)=\alpha_+(\lambda)
 \bigl(1+\gamma F(\lambda)\bigr),\qquad
 \lambda\in\mathbb{R}.
 \end{equation}
The limiting values $\alpha_\pm(\lambda)$ can be continued  to the
upper (resp. to the lower) half-plane, where they are non-vanishing.
At $\lambda\to\infty$ the function $\alpha(\lambda)$ behaves as
 \begin{equation}\label{As-al}
 \alpha(\lambda)=1+\frac{\alpha_1}\lambda+O(\lambda^{-2}), \qquad
 \lambda\to\infty, \qquad\mbox{where}\qquad \alpha_1=-\int\limits_{\mathbb{R}}\nu(\mu)\,d\mu.
 \end{equation}

Let us also define $(\np\times \nm)$-matrix $A^-$ and $(\nm\times
\np)$-matrix $A^+$ as
 \begin{equation}\label{A-A+A}
 A^-_{jk}=\frac{h^-_ke_-^2(\qm_k)}{\qp_j-\qm_k},
 \qquad
  A^+_{jk}=\frac{h^+_ke_+^2(\qp_k)}{\qm_j-\qp_k}, \qquad
  \mbox{where}\quad h^\pm_k=-\frac{\bigl(\alpha_\pm(\qpm_k)\bigr)^{\mp2}}{\gamma
 F'(\qpm_k)}.
  \end{equation}

We also introduce a set of contours $\Gamma^{(n)}_{J;K}$ with
$n=0,1,\dots,\min(\np,\nm)$. Here $J$ and $K$ are multi-indexes:
$J=\{j_1,\dots,j_n\}$ with $1\le j_s\le \np$, and
$K=\{k_1,\dots,k_n\}$ with $1\le k_s\le \nm$. We set by definition
$\Gamma^{(0)}=\mathbb{R}$. The contour $\Gamma^{(n)}_{J;K}$ is a
deformation of the real axis such that moving $\mathbb{R}$ to
$\Gamma^{(n)}_{J;K}$ we cross only the roots $\qp_{j_1},\dots,
\qp_{j_n}$ and $\qm_{k_1},\dots, \qm_{k_n}$, while other roots
$\qpm_\ell$ and all the poles $\Rpm_\ell$ should not be crossed (see
Fig.~\ref{G1212}).

Finally let
 \begin{equation}\label{A-rep}
 {\cal A}_{\cal C}([g],[\nu])=-\int\limits_{\cal C}
 \bigl(ix+g'(\lambda)\bigr)\nu(\lambda)\,d\lambda +\int\limits_{\cal C}
 \frac{\nu(\lambda)\nu(\mu)}{(\lambda-\mu_+)^2}\,d\lambda\,d\mu.
 \end{equation}
Here the contour ${\cal C}$ is one of $\Gamma^{(n)}_{J;K}$. The
symbol $\mu_+$ means that $\mu$ is slightly shifted to the left from
the integration contour ${\cal C}$. Note that one can always choose
the cuts of the function $\nu(\lambda)$ in such a way that the
contour ${\cal C}$ does not cross them.


Now we are ready to formulate the main theorems on the asymptotic
behavior of the resolvent and ${\det}_{\mathbb{R}}(I+\gamma V)$.

\begin{thm}\label{Main-thm-Res}
Let $x\to\infty$ and
 \begin{equation}\label{def-Res}
 R(\lambda,\mu)+\gamma\int\limits_{\mathbb{R}}V(\lambda,\xi)
 R(\xi,\mu)\,d\xi = V(\lambda,\mu).
 \end{equation}
Then $R(\lambda,\mu)$ has the following form:
 \begin{equation}
 R(\lambda,\mu)=\frac{\sqrt{F(\lambda)F(\mu)} }{2i\pi (\lambda-\mu)}
  \bigl[ f_+(\lambda)f_-(\mu)-f_-(\lambda)f_+(\mu) \bigr], \quad
 \label{Res-GSK}
\end{equation}
where
 \begin{equation}\label{ANS-f+}
 f_\pm(\lambda)=\alpha_\mp^{\mp1}(\lambda)e_\pm(\lambda)\left[1+\sum_{j=1}^{\nmp}
 \frac{D^\pm_jh_j^\mp e_\mp^2(\qmp_j)}{\lambda-\qmp_j}\right]
 + \alpha_\pm^{\pm1}(\lambda)e_\mp(\lambda)\sum_{j=1}^{\npm}
 \frac{C^\pm_jh_j^\pm e_\pm^2(\qpm_j)}{\lambda-\qpm_j}+O\left(e^{-ax}\right),
\end{equation}
uniformly for $\lambda\in\mathbb{R}$. The constants $\Cpm_j$ and $\Dpm_j$
can be found from the systems
 \begin{equation}\label{Sys-CD}
 \left\{
 \begin{array}{l}
 \Cp_j-\sum_{k=1}^{\np}A_{jk}\Cp_k=1,\\
 \Dp_j-\sum_{k=1}^{\np}A_{jk}^+\Cp_k=0,
 \end{array}\right. \qquad\qquad
 \left\{
 \begin{array}{l}
 \Cm_j-\sum_{k=1}^{\nm}\tilde A_{jk}C^-_k=1,\\
 \Dm_j-\sum_{k=1}^{\nm}A_{jk}^-C^-_k=0,
 \end{array}\right.
 \end{equation}
where $A=A^-A^+$ and $\tilde A=A^+A^-$.
\end{thm}

This theorem is proved in the next section.

{\sl Remark.} The entries of the matrices $A^\pm$ are exponentially
small for $x$ large enough. Therefore $\det(I-A)\ne 0$ and
$\det(I-\tilde A)\ne 0$, and hence, each of the systems
\eqref{Sys-CD} has a unique solution.

\begin{thm}\label{Main-thm-FD}
Let $x\to\infty$. Then the Fredholm determinant of the operator
$I+\gamma V$ behaves as
 \begin{equation}\label{logdet1}
 {\det}_{\mathbb{R}}(I+\gamma V)=e^{{\cal A}_{\mathbb{R}}([g],[\nu])}
 \det_{~~\np}(I-A)\left(1+O\left(e^{-ax}\right)\right).
 \end{equation}
Here the functional  ${\cal A}_{\mathbb{R}}([g],[\nu])$ is given by
\eqref{A-rep} with ${\cal C}=\mathbb{R}$.
\end{thm}

The proof of this theorem is given in section~\ref{S-Main-thm-FD}.

{\sl Remark.} In \eqref{logdet1} the determinant of $(\np\times
\np)$-matrix $I-A$ can be replaced by the determinant of $(\nm\times
\nm)$-matrix $I-\tilde A$.

\begin{thm}\label{Main-thm-FD1} Let $x\to\infty$. Then the Fredholm determinant
of the operator $I+\gamma V$ behaves as
 \begin{equation}\label{logdet0}
 {\det}_{\mathbb{R}}(I+\gamma V)=\sum_{{\cal C}}
  e^{{\cal A}_{\cal C}([g],[\nu])}
 \left(1+O\left(e^{-ax}\right)\right),
 \end{equation}
where the functional  ${\cal A}_{\cal C}([g],[\nu])$ is given by
\eqref{A-rep} and the sum is taken with respect to all possible
contours ${\cal C}\in\{\Gamma^{(n)}_{J;K}\}$ including
$\Gamma^{(0)}=\mathbb{R}$.
\end{thm}

The proof of this theorem is given in section~\ref{S-Main-thm-FD1}.

\section{Asymptotic formula for the resolvent\label{Asy-form-Res}}

This section is devoted to the proof of Theorem~\ref{Main-thm-Res}.
The operator $I+\gamma V$ belongs to the class of completely
integrable operators \cite{IIKS90,Dei99}. It is known
\cite{ItsIK90,IIKS90,Dei99} that for such operators the kernel of
the resolvent has the form \eqref{Res-GSK}, where functions
$f_\pm(\lambda)$ solve an integral equation
 \begin{equation}\label{Int-eqF+-}
  f_\pm(\lambda)+\frac{\gamma}{2\pi i} \int\limits_{\mathbb{R}}
 \frac{e_+(\lambda)e_-(\mu)-e_-(\lambda)e_+(\mu)}{\lambda-\mu}
 F(\mu)f_\pm(\mu)\,d\mu=e_\pm(\lambda).
 \end{equation}
Thus, to prove Theorem~\ref{Main-thm-Res} we should solve
asymptotically the equation \eqref{Int-eqF+-} up to
$O\left(e^{-ax}\right)$ terms.

{\sl Proof of Theorem~\ref{Main-thm-Res}.} Since for  $\gamma$ small
enough ${\det}_{\mathbb{R}}(I+\gamma V)\ne 0$,  the solution of
\eqref{Int-eqF+-} exits and unique. Therefore it is enough to
substitute \eqref{ANS-f+} into the integral equation
\eqref{Int-eqF+-} and to check that the last one holds up to terms
of order $O\left(e^{-ax}\right)$.

Consider, for instance, the equation for $f_+(\lambda)$. Making the substitution we
find
 \begin{multline}\label{f+}
 f_+(\lambda)- e_+(\lambda)=\frac1{2\pi i}\int\limits_{\mathbb{R}}\frac{\gamma
 F(\mu)\alpha_-^{-1}(\mu)\,d\mu}{\mu-\lambda+i0}\left[1+\sum_{j=1}^{\nm}
 \frac{\Dp_jh_j^-e_-^2(\qm_j)}{\mu-\qm_j}\right]\bigl
 (e_+(\lambda)-e_-(\lambda)e_+^2(\mu)\bigr)\numa{30}
 +\frac1{2\pi i}\int\limits_{\mathbb{R}}\frac{\gamma
 F(\mu)\alpha_+(\mu)\,d\mu}{\mu-\lambda-i0}\sum_{j=1}^{\np}
 \frac{\Cp_jh_j^+e_+^2(\qp_j)}{\mu-\qp_j}\bigl
 (e_+(\lambda)e_-^2(\mu)-e_-(\lambda)\bigr)+O\left(e^{-ax}\right).
 \end{multline}
For convenience we have shifted $\mu-\lambda$ by $+i0$ in the first
integral and by $-i0$ in the second one. Consider the
coefficient at $e_+(\lambda)$.  We have
 \begin{multline}\label{Ce+}
 \alpha_-^{-1}(\lambda)\left[1+\sum_{j=1}^{\nm}
 \frac{\Dp_jh_j^-e_-^2(\qm_j)}{\lambda-\qm_j}\right]-1=
 \frac1{2\pi i}\int\limits_{\mathbb{R}}\frac{\gamma
 F(\mu)\alpha_-^{-1}(\mu)\,d\mu}{\mu-\lambda+i0}\left[1+\sum_{j=1}^{\nm}
 \frac{\Dp_jh_j^-e_-^2(\qm_j)}{\mu-\qm_j}\right]\numa{30}
 + \frac1{2\pi i}\int\limits_{\mathbb{R}}\frac{\gamma
 F(\mu)\alpha_+(\mu)e_-^2(\mu)\,d\mu}{\mu-\lambda-i0}\sum_{j=1}^{\np}
 \frac{\Cp_jh_j^+e_+^2(\qp_j)}{\mu-\qp_j}+O\left(e^{-ax}\right).
 \end{multline}
The integral in the first line of \eqref{Ce+} can be taken
explicitly by use of  $\gamma
F(\lambda)\alpha_-^{-1}(\lambda)=\alpha_+^{-1}(\lambda)
-\alpha_-^{-1}(\lambda)$. Since $\alpha_\pm(\lambda)$ are analytical
and non-vanishing in the corresponding half-planes and due to the
condition \eqref{As-al} we obtain,
 \begin{multline}\label{1-int}
  \frac1{2\pi i}\int\limits_{\mathbb{R}}\frac{\gamma
 F(\mu)\alpha_-^{-1}(\mu)\,d\mu}{\mu-\lambda+i0}\left[1+\sum_{j=1}^{\nm}
 \frac{\Dp_jh_j^-e_-^2(\qm_j)}{\mu-\qm_j}\right]\numa{30}
 =\frac1{2\pi i}\int\limits_{\mathbb{R}}\,d\mu\frac{\alpha_+^{-1}(\mu)-1-(\alpha_-^{-1}(\mu)-1)}{\mu-\lambda+i0}
 +\frac1{2\pi i}\int\limits_{\mathbb{R}}\,d\mu\frac{\alpha_+^{-1}(\mu)-\alpha_-^{-1}(\mu)}{\mu-\lambda+i0}\sum_{j=1}^{\nm}
 \frac{\Dp_jh_j^-e_-^2(\qm_j)}{\mu-\qm_j}\numa{30}
 =\alpha_-^{-1}(\lambda)\left[1+\sum_{j=1}^{\nm}
 \frac{\Dp_jh_j^-e_-^2(\qm_j)}{\lambda-\qm_j}\right]-1-\sum_{j=1}^{\nm}
 \frac{\alpha_-^{-1}(\qm_j)\Dp_jh_j^-e_-^2(\qm_j)}{\lambda-\qm_j}.
 \end{multline}
Thus, the equation \eqref{Ce+} takes the form
 \begin{equation}\label{Ce+1}
 \sum_{j=1}^{\nm} \frac{\alpha_-^{-1}(\qm_j)\Dp_jh_j^-e_-^2(\qm_j)}{\lambda-\qm_j}=
 \frac1{2\pi i}\int\limits_{\mathbb{R}}\frac{\gamma F(\mu)\alpha_-(\mu)e_-^2(\mu)\,d\mu}{\bigl(1+\gamma
F(\mu)\bigr)(\mu-\lambda-i0)}\sum_{k=1}^{\np}
 \frac{\Cp_kh_k^+e_+^2(\qp_k)}{\mu-\qp_k}+O\left(e^{-ax}\right).
 \end{equation}
Here we have used $\alpha_+(\mu)=\alpha_-(\mu)\bigl(1+\gamma
F(\mu)\bigr)^{-1}$. The remaining integral can be computed
asymptotically by the residues at $1+\gamma F(\mu)=0$ in the lower
half-plane. Comparing then the coefficients at every
$(\lambda-\qm_j)^{-1}$ we obtain
 \begin{equation}\label{D-C}
 \Dp_j=\sum_{k=1}^{\np}A_{jk}^+\Cp_k.
 \end{equation}

Similar calculation of the coefficient at $e_-(\lambda)$ in
\eqref{f+} leads us to equations
 \begin{equation}\label{C-D}
 \Cp_j=1+\sum_{k=1}^{\nm}A_{jk}^-\Dp_k.
 \end{equation}
The equations \eqref{D-C}, \eqref{C-D} yield immediately the first
of the systems \eqref{Sys-CD}. The second system \eqref{Sys-CD}
follows from the analysis of the integral equation for the function
$f_-(\lambda)$, what can be done by the same method. \qed

\section{The first asymptotic formula for the determinant\label{S-Main-thm-FD}}

The leading terms  of the Fredholm determinant
$\det_{\mathbb{R}}(I+\gamma V)$ asymptotic expansion were obtained
in the work \cite{KitKMST09b} (see also \cite{Akh64} for the case
$g(\lambda)=0$)
 \begin{equation}\label{Lead-Asy-FD}
 \log{\det}_{\mathbb{R}}(I+\gamma V)=
 {\cal A}_{\mathbb{R}}([g],[\nu])+o(1),\qquad x\to\infty.
 \end{equation}
Thus, we need only to prove that the corrections to this formula have the form of the finite size matrix determinant $\det(I-A)$.

{\sl Proof of Theorem~\ref{Main-thm-FD}.} The corrections to the
equation \eqref{Lead-Asy-FD} can be computed from the following
identity \cite{ItsIK90,IIKS90}:
 \begin{equation}\label{x-der}
 \partial_x\log{\det}_{\mathbb{R}}(I+\gamma V)=\frac\gamma{2\pi}
 \int\limits_{\mathbb{R}}f_+(\lambda)e_-(\lambda)F(\lambda)\,d\lambda.
 \end{equation}
Substituting here  \eqref{ANS-f+} for $f_+(\lambda)$  we obtain
 \begin{multline}\label{x-der1}
 \partial_x\log\det(I+\gamma V)=\frac1{2\pi}\int\limits_{\mathbb{R}}\gamma F(\lambda)\alpha_-^{-1}(\lambda)
 \left[1+\sum_{j=1}^{\nm}
 \frac{\Dp_jh_j^-e_-^2(\qm_j)}{\lambda-\qm_j}\right]\,d\lambda\numa{30}
 +
 \frac1{2\pi}\int\limits_{\mathbb{R}}
 \gamma F(\lambda)\alpha_-(\lambda)e_-^2(\lambda)\sum_{j=1}^{\np}
 \frac{\Cp_jh_j^+e_+^2(\qp_j)}{\lambda-\qp_j}+O\left(e^{-ax}\right).
\end{multline}
Let us give several comments on the calculation of
the integral
 \begin{equation}\label{1int}
 \frac1{2\pi}\int\limits_{\mathbb{R}}\gamma F(\lambda)\alpha_-^{-1}(\lambda)
 \,d\lambda.
 \end{equation}
We have
 \begin{equation}\label{a-a}
 \gamma F(\lambda)\alpha_-^{-1}(\lambda)=\alpha_+^{-1}(\lambda)-\alpha_-^{-1}(\lambda)
 =\left(\alpha_+^{-1}(\lambda)-1+\frac{\alpha_1}{\lambda-\lambda_0}\right)
 -\left(\alpha_-^{-1}(\lambda)-1+\frac{\alpha_1}{\lambda-\lambda_0}\right),
 \end{equation}
where $\lambda_0$ is an arbitrary complex number with positive
imaginary part. Observe that due to \eqref{As-al} both terms in the
r.h.s. of \eqref{a-a} behave as $O(\lambda^{-2})$ as
$\lambda\to\infty$. Therefore one can integrate each of these terms
separately. The integral of the second term vanishes, since the
integrand is analytical in the lower half-plane. The first term in
\eqref{a-a} has only one simple pole in the upper half-plane,
therefore
 \begin{equation}\label{2int}
 \frac1{2\pi}\int\limits_{\mathbb{R}}\gamma F(\lambda)\alpha_-^{-1}(\lambda)
 \,d\lambda=\frac1{2\pi}\int\limits_{\mathbb{R}}
 \left(\alpha_+^{-1}(\lambda)-1+\frac{\alpha_1}{\lambda-\lambda_0}\right)
 \,d\lambda=i\alpha_1.
 \end{equation}
Other integrals in \eqref{x-der1} can be computed similarly to the
ones considered in the previous section. Using the equations
\eqref{Sys-CD} we obtain after simple algebra
 \begin{equation}\label{x-der3}
 \partial_x\log\det(I+\gamma V)=i\alpha_1-\sum_{j=1}^{\np} A'_{jj}\Cp_j+O\left(e^{-ax}\right),
 \end{equation}
where prime means the derivative over $x$. Using the explicit
expression \eqref{A-A+A} for $A$ one can easily convince himself
that
 \begin{equation}\label{ident-A}
 A'_{jj}=A'_{\ell j}+i(\qp_\ell-\qp_j)A_{\ell j}.
 \end{equation}
Then due to \eqref{Sys-CD} we have
 \begin{multline}\label{x-der3-1}
 \partial_x\log\det(I+\gamma V)-i\alpha_1=-\sum_{j,\ell=1}^{\np}\left[
 A'_{\ell j}+i(\qp_\ell-\qp_j)A_{\ell j}\right](I-A)^{-1}_{j\ell}+O\left(e^{-ax}\right)\\
 = \partial_x\tr\log(I-A)+i\sum_{j,\ell=1}^{\np}(\qp_\ell-\qp_j)\bigl(\delta_{\ell j}-A_{\ell
 j}\bigr)(I-A)^{-1}_{j\ell}+O\left(e^{-ax}\right)\\
 =\partial_x\log\det(I-A)-i\sum_{j=1}^{\np}\qp_j+i\sum_{\ell=1}^{\np}\qp_\ell+O\left(e^{-ax}\right)
 =\partial_x\log\det(I-A)+O\left(e^{-ax}\right).
 \end{multline}
Integrating over $x$ we arrive at
 \begin{equation}\label{logdet11}
 \log\det(I+\gamma V)=ix\alpha_1+C+\log\det(I-A)+O\left(e^{-ax}\right),
 \end{equation}
and comparing \eqref{logdet11} with \eqref{Lead-Asy-FD} we find the
integration constant $C$
 \begin{equation}\label{A-rep-1}
 C=-\int\limits_{\mathbb{R}}g'(\lambda)\nu(\lambda)\,d\lambda+
 \int\limits_{\mathbb{R}}\frac{\nu(\lambda)\nu(\mu)}{(\lambda-\mu_+)^2}\,d\lambda\,d\mu=
  \Bigl.{\cal A}_{\mathbb{R}}([g],[\nu])\Bigr|_{x=0}.
 \end{equation}
Thus, we have found the corrections to \eqref{Lead-Asy-FD}, and it
remains to prove that the determinant of $(\np\times\np)$-matrix
$I-A$ is equal to the determinant of $(\nm\times\nm)$-matrix
$I-\tilde A$. For this it is enough to present $\det(I-A)$ as a
determinant of a block-matrix of the size
$(\np+\nm)\times(\np+\nm)$:
 \begin{equation}\label{detA}
 \det(I-A)=\det(I-A^-A^+)=\det\left(
 \begin{array}{cc}I&A^+\\A^-&I\end{array}\right)=\det(I-A^+A^-)
 =\det(I-\tilde A),
 \end{equation}
what ends the proof. \qed

\section{The second asymptotic formula for the determinant\label{S-Main-thm-FD1}}

Define  a set of analogs of the function $\alpha(\lambda)$
\eqref{nu-1}
 \begin{equation}\label{alpha_Ga}
 \alpha\bigl(\lambda;\Gamma^{(n)}_{J;K}\bigr)=
 \exp\biggl(\hspace{1mm}\int\limits_{\Gamma^{(n)}_{J;K}}
 \frac{\nu(\mu)\,d\mu}{\mu-\lambda}\biggr),
 \end{equation}
where $\nu(\lambda)$ is still given by \eqref{nu-1}. Obviously,
 \begin{equation}\label{fact-a}
 1+\gamma F(\lambda)=\frac{\alpha_-\bigl(\lambda;\Gamma^{(n)}_{J;K}\bigr)}
 {\alpha_+\bigl(\lambda;\Gamma^{(n)}_{J;K}\bigr)},  \qquad
 \lambda\in\Gamma^{(n)}_{J;K}.
 \end{equation}
Here $\alpha_\pm$ are the limiting values of the function
$\alpha\bigl(\lambda;\Gamma^{(n)}_{J;K}\bigr)$ on the contour
$\Gamma^{(n)}_{J;K}$ from the left (resp. from the right). The
function $\alpha_-$ is analytical and non-vanishing in the domain to
the right from $\Gamma^{(n)}_{J;K}$, it has zeros at
$\lambda=\qp_\ell$, $\ell\notin J$ and poles at $\lambda=\Rp_\ell$,
$\ell=1,\dots,\np$. Similarly  $\alpha_+^{-1}$ is analytical and
non-vanishing in the domain to the left from $\Gamma^{(n)}_{J;K}$,
it has zeros at $\lambda=\qm_\ell$, $\ell\notin K$ and poles at
$\lambda=\Rm_\ell$, $\ell=1,\dots,\nm$.

\begin{lemma}\label{Lem-cont} Let $\nu(\lambda)$ be given by \eqref{nu-1}.
Then
\begin{equation}\label{rat}
 \exp\left\{{\cal A}_{\Gamma^{(n)}_{J;K}}([g],[\nu])
 -{\cal A}_{\Gamma^{(n-1)}_{\hat J;\hat K}}([g],[\nu])\right\}
 =\frac{\alpha^2_-\bigl(\qm_n;\Gamma^{(n-1)}_{\hat J;\hat K}\bigr)
 \alpha^{-2}_+\bigl(\qp_n;\Gamma^{(n-1)}_{\hat J;\hat K}\bigr)}
 {(\qp_{j_n}-\qm_{k_n})^2\gamma^2
 F'(\qp_{j_n})F'(\qm_{k_n})}e_+^2(\qp_{j_n})e_-^2(\qm_{k_n}),
 \end{equation}
where $\hat J=J\setminus j_n$ and $\hat K=K\setminus k_n$.
\end{lemma}

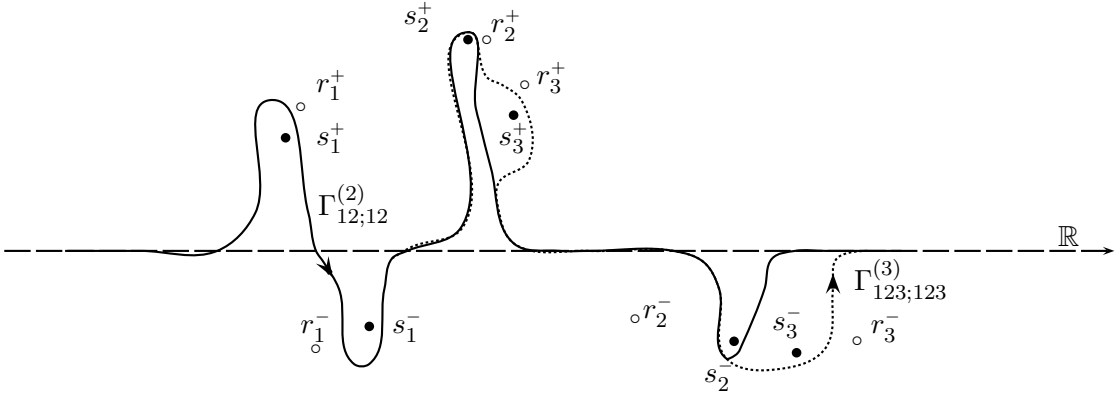
\begin{figure}[h]
\begin{center}
 \begin{pspicture}(15,5)
 \psline[linestyle=dashed, dash=10pt 2pt]{->}(0.2,2)(14.8,2)
 \pscurve[linestyle=dashed, dash=1pt 1pt]
 (5.3,1.9)(5.55,2.05)(6.22,2.3)
 (6.3,4.9)(6.55,4.3)(6.75,4.2)(7,4)(7.1,3.2)(6.7,2.9)(6.7,2.55)(6.9,2.1)
 (7.9,2)(9,2)(9.4,1.85)(9.6,1.5)(9.7,0.57)(10.5,0.45)(11,0.71)(11.1,1.5)(11.15,1.9)(11.55,2)
 \pscurve(1,2)(2,2)(3,2)(3.5,2.5)(3.7,4)(4.2,2.5)(4.3,2)(4.4,1.85)(4.6,1.5)(4.8,0.5)(5,0.5)(5.2,1.5)
 (5.3,1.9)(5.5,2)(6.2,2.3)(6.3,4.9)(6.4,3.9)(6.45,3.7)(6.6,3.1)(6.9,2.1)(7.5,2)
 (7.9,2)(9,2)(9.4,1.85)(9.6,1.5)(9.7,0.57)(9.75,0.58)(9.85,0.65)(9.9,0.75)(10.2,1.5)(10.3,1.9)(10.6,2)(11,2)(12,2)
 \rput(4.5,4.2){$\Rp_1$}%
 \rput(4.1,3.9){$\scriptstyle\circ$}
 \rput(4.5,3.5){$\qp_1$}%
 \psdots(3.9,3.5)
 \rput(5.5,1){$\qm_1$}
 \rput(4.3,1){$\Rm_1$}
 \rput(4.3,0.7){$\scriptstyle\circ$}%
 \psdots(5,1)
 \rput(4.8,2.6){$\Gamma^{(2)}_{12;12}$}
  \rput(12,1.6){$\Gamma^{(3)}_{123;123}$}
 \rput(14.2,2.2){$\mathbb{R}$}
 \rput(7.4,4.3){$\Rp_3$}%
 \rput(7.05,4.2){$\scriptstyle\circ$}
 \rput(5.65,5.1){$\qp_2$}%
 \psdots(6.3,4.8)
 \rput(6.9,3.5){$\qp_3$}
 \rput(6.8,4.99){$\Rp_2$}
 \rput(6.55,4.8){$\scriptstyle\circ$}%
 \psdots(6.9,3.8)
 \rput(8.8,1.2){$\Rm_2$}%
 \rput(8.5,1.1){$\scriptstyle\circ$}
 \rput(9.6,0.35){$\qm_2$}%
 \psdots(9.8,0.8)
 \rput(10.5,1.05){$\qm_3$}
 \rput(11.8,1){$\Rm_3$}
 \rput(11.42,0.8){$\scriptstyle\circ$}%
 \psdots(10.62,0.65)
 %
 \psline[linewidth=2pt]{<-}(4.52,1.65)(4.49,1.7)
 \psline[linewidth=2pt]{<-}(11.1,1.7)(11.1,1.5)
\end{pspicture}
\caption{\label{G1212-G123123}Integration contours
$\Gamma^{(2)}_{12;12}$ (solid line) and $\Gamma^{(3)}_{123;123}$
(dotted line)}
\end{center}
\end{figure}

{\sl Proof.} Consider the first term in \eqref{A-rep} setting ${\cal
C}=\Gamma^{(n)}_{J;K}$ and ${\cal C}=\Gamma^{(n-1)}_{\hat J;\hat
K}$. The contour $\Gamma^{(n)}_{J;K}$ is obtained from
$\Gamma^{(n-1)}_{\hat J;\hat K}$ by crossing the roots of $1+\gamma
F(\lambda)$ at $\qp_{j_n}$ and $\qm_{k_n}$ (see
Fig~\ref{G1212-G123123}). Therefore we have

%
%
 \begin{equation}\label{diff-sing-int}
-\biggl(\int\limits_{\Gamma^{(n)}_{J;K}}-\int\limits_{\Gamma^{(n-1)}_{\hat
J;\hat K}}\biggr)
 \bigl(ix+g'(\lambda)\bigr)\nu(\lambda)\,d\lambda=
 \int\limits_{\qm_{k_n}}^{\qp_{j_n}}
 \bigl(ix+g'(\lambda)\bigr)\,d\lambda=\log\left(e_+^2(\qp_{j_n})e_-^2(\qm_{k_n})\right).
 \end{equation}

Let us transform now the double integral in  \eqref{A-rep}. To
lighten the notations we set
 \begin{equation}\label{light-not}
 \alpha\bigl(\lambda;\Gamma^{(n)}_{J;K}\bigr)=
 \alpha\bigl(\lambda;n), \qquad
 \alpha\bigl(\lambda;\Gamma^{(n-1)}_{\hat J;\hat K}\bigr)=
 \alpha\bigl(\lambda;n-1).
 \end{equation}
We can present $\nu(\lambda)$ on the contour $\Gamma^{(n-1)}_{\hat J;\hat K}$ as
 \begin{equation}\label{fact-a1}
 \nu(\lambda)=\frac{-1}{2\pi i}
 \log\frac{\alpha_-(\lambda;n-1)}
 {\alpha_+(\lambda;n-1)}, \qquad
 \lambda\in\Gamma^{(n-1)}_{\hat J;\hat K}.
 \end{equation}
Using this representation one can integrate over $\lambda$ in the
double integral in \eqref{A-rep}
 \begin{equation}\label{int-ov-l1}
 \int\limits_{\Gamma^{(n-1)}_{\hat J;\hat K}}
 \frac{\nu(\lambda)\nu(\mu)}{(\lambda-\mu_+)^2}\,d\lambda\,d\mu=
 \int\limits_{\Gamma^{(n-1)}_{\hat J;\hat K}}\nu(\mu)\cdot
 \log'\alpha_+(\mu;n-1)\,d\mu.
 \end{equation}
Substituting here $\nu(\mu)$ via \eqref{fact-a1} we arrive at
 \begin{equation}\label{int-ov-l2}
 \int\limits_{\Gamma^{(n-1)}_{\hat J;\hat K}}
 \frac{\nu(\lambda)\nu(\mu)}{(\lambda-\mu_+)^2}\,d\lambda\,d\mu=
 \frac{-1}{2\pi i}\int\limits_{\Gamma^{(n-1)}_{\hat J;\hat K}}
 \log\alpha_-(\mu;n-1)\cdot
 \log'\alpha_+(\mu;n-1)\,d\mu.
 \end{equation}
Similarly we have on the contour $\Gamma^{(n)}_{J;K}$
 \begin{equation}\label{int-ov-l3}
 \int\limits_{\Gamma^{(n)}_{J;K}}
 \frac{\nu(\lambda)\nu(\mu)}{(\lambda-\mu_+)^2}\,d\lambda\,d\mu=
 \frac{-1}{2\pi i}\int\limits_{\Gamma^{(n)}_{J;K}}
 \log\alpha_-(\mu;n)\cdot
 \log'\alpha_+(\mu;n)\,d\mu.
 \end{equation}

In order to compare \eqref{int-ov-l2} and \eqref{int-ov-l3} we first
observe that
 \begin{equation}\label{ar-ag}
 \alpha_\pm(\lambda;n)=\alpha_\pm(
 \lambda;n-1)
 \left(\frac{\lambda-\qm_{k_n}}{\lambda-\qp_{j_n}}\right).
 \end{equation}
Let us  introduce now $\hat\alpha_\pm(\lambda)$ by
 \begin{equation}\label{ha}
 \alpha_-(\lambda;n-1)=
 \hat\alpha_-(\lambda)\left(\frac{\lambda-\qp_{j_n}}{\lambda-\Rp_{j_n}}\right),\qquad
 \alpha_+(\lambda;n-1)=\hat\alpha_+(\lambda)\left(\frac{\lambda-\Rm_{k_n}}
 {\lambda-\qm_{k_n}}\right).
 \end{equation}
Due to \eqref{ar-ag} we also have
 \begin{equation}\label{ha1}
  \alpha_-(\lambda;n)=\hat\alpha_-(\lambda)\left(\frac{\lambda-\qm_{k_n}}
  {\lambda-\Rp_{j_n}}\right),\qquad
 \alpha_+(\lambda;n)=\hat\alpha_+(\lambda)\left(\frac{\lambda-\Rm_{k_n}}
 {\lambda-\qp_{j_n}}\right).
 \end{equation}
It is easy to see that $\hat\alpha_-(\lambda)$ is analytical and
non-vanishing to the right from $\Gamma^{(n-1)}_{\hat J;\hat K}$ and
$\Gamma^{(n)}_{J;K}$, while $\hat\alpha_+^{-1}(\lambda)$ is
analytical and non-vanishing to the left from $\Gamma^{(n-1)}_{\hat
J;\hat K}$ and $\Gamma^{(n)}_{J;K}$. Moreover, both
$\hat\alpha_\pm(\lambda)$ are analytical and non-vanishing in the
interior between $\Gamma^{(n-1)}_{\hat J;\hat K}$ and
$\Gamma^{(n)}_{J;K}$.

Substituting \eqref{ha}, \eqref{ha1} into \eqref{int-ov-l2} and
\eqref{int-ov-l3} we obtain
 \begin{equation}\label{C1R-ha}
 \int\limits_{\Gamma^{(n-1)}_{\hat J;\hat K}}
 \frac{\nu(\lambda)\nu(\mu)}{(\lambda-\mu_+)^2}\,d\lambda\,d\mu
 =\frac{-1}{2\pi i}
 \int\limits_{\Gamma^{(n-1)}_{\hat J;\hat K}}\log'\hat\alpha_+(\mu)\cdot\log\hat\alpha_-(\mu)\,d\mu+
 \log\left(\frac{\alpha_-(\Rm_{k_n};n-1)\hat\alpha_+(\qp_{j_n})}{\alpha_-(\qm_{k_n};n-1)\hat\alpha_+(\Rp_{j_n})}\right),
 \end{equation}
and
 \begin{equation}\label{C1G-ha}
 \int\limits_{\Gamma^{(n)}_{J;K}}
 \frac{\nu(\lambda)\nu(\mu)}{(\lambda-\mu_+)^2}\,d\lambda\,d\mu=
 \frac{-1}{2\pi i}
 \int\limits_{\Gamma^{(n)}_{J;K}}\log'\hat\alpha_+(\mu)\cdot\log\hat\alpha_-(\mu)\,d\mu
 + \log\left(\frac{\alpha_-(\Rm_{k_n};n)
 \hat\alpha_+(\qm_{k_n})}
 {\alpha_-(\qp_{j_n};n)\hat\alpha_+(\Rp_{j_n})}\right).
 \end{equation}
Using that $\log\hat\alpha_\pm(\mu)$ is holomorphic in the interior
between $\Gamma^{(n-1)}_{\hat J;\hat K}$ and $\Gamma^{(n)}_{J;K}$ we
find
 \begin{equation}\label{JGR-prem}
 \biggl(\;\int\limits_{\Gamma^{(n)}_{J;K}}-\int\limits_{\Gamma^{(n-1)}_{\hat J;\hat K}}\biggr)
 \frac{\nu(\lambda)\nu(\mu)}{(\lambda-\mu_+)^2}\,d\lambda\,d\mu=\log\left(
 \frac{\alpha_-(\qm_{k_n};n-1)
 \cdot\alpha_-(\Rm_{k_n};n)\cdot\hat\alpha_+(\qm_{k_n})}
 {\alpha_-(\Rm_{k_n};n-1)\cdot\alpha_-(\qp_{j_n};n)\cdot\hat\alpha_+(\qp_{j_n})}
 \right).
 \end{equation}

The last step is to express the obtained answer in terms of
$\alpha_\pm(\lambda;n-1)$ via \eqref{ar-ag}, \eqref{ha}. We will
also need the following evident equations:
 \begin{equation}\label{der1}
 \begin{array}{l}
 \hat\alpha_-(\qp_{j_n})=(\qp_{j_n}-\Rp_{j_n})\bigl.
 \frac{d}{d\lambda}\alpha_-(\lambda;n-1)\bigr|_{\lambda=\qp_{j_n}},\num
 \hat\alpha_+^{-1}(\qm_{k_n})=(\qm_{k_n}-\Rm_{k_n})\bigl.
 \frac{d}{d\lambda}\alpha_+^{-1}(\lambda;n-1)\bigr|_{\lambda=\qm_{k_n}},
  \end{array}
 \end{equation}
and
 \begin{equation}\label{der2}
 \begin{array}{l}
 \gamma F'(\qp_{j_n})=
 \alpha_+^{-1}\bigl(\qp_{j_n};n-1)\cdot
 \bigl.\frac{d}{d\lambda}\alpha_-(\lambda;n-1)\bigr|_{\lambda=\qp_{j_n}},\num
 \gamma
 F'(\qm_{k_n})=\alpha_-(\qm_{k_n};n-1)
 \bigl.\frac{d}{d\lambda}\alpha_+^{-1}(\lambda;n-1)\bigr|_{\lambda=\qm_{k_n}}.
 \end{array}
 \end{equation}
Substituting these formulas into \eqref{JGR-prem} we obtain after
simple algebra
 \begin{equation}\label{JGR-fin}
 \biggl(\;\int\limits_{\Gamma^{(n)}_{J;K}}-\int\limits_{\Gamma^{(n-1)}_{\hat J;\hat K}}\biggr)
 \frac{\nu(\lambda)\nu(\mu)}{(\lambda-\mu_+)^2}\,d\lambda\,d\mu=
 \log\left(\frac{\alpha_-^2(\qm_{k_n};n-1)
 \alpha_+^{-2}(\qp_{j_n};n-1)}{(\qp_{j_n}-\qm_{k_n})^2\gamma^2 F'(\qp_{j_n})F'(\qm_{k_n})}\right).
 \end{equation}
Combining this result with \eqref{diff-sing-int} we arrive at the
statement of Lemma.\qed

\begin{cor}\label{An-A0}
Let $\alpha(\lambda)$ be defined by \eqref{nu-1}. Then
 \begin{equation}\label{JGR-fin1}
  \exp\left\{{\cal A}_{\Gamma^{(n)}_{J;K}}([g],[\nu])-{\cal A}_{\mathbb{R}}([g],[\nu])\right\}=
  \left(\det\frac1{\qp_{j_a}-\qm_{k_b}}\right)^2
  \prod_{m=1}^n\left(\frac{\alpha_-(\qm_{k_m})}
 {\alpha_+(\qp_{j_m})}\right)^2\frac{e_+^2(\qp_{j_m})e_-^2(\qm_{k_m})}{\gamma^2
 F'(\qp_{j_m})F'(\qm_{k_m})}.
 \end{equation}
\end{cor}

{\sl Proof.} Applying successively Lemma~\ref{Lem-cont} to the
exponents
$$
\exp\left\{{\cal
A}_{\Gamma^{(m)}_{j_1,\dots,j_m;k_1,\dots,k_m}}([g],[\nu])-{\cal
A}_{\Gamma^{(m-1)}_{j_1,\dots,j_{m-1};k_1,\dots,k_{m-1}}}([g],[\nu])\right\}
$$
for $m=1,\dots,n$ we have
 \begin{equation}\label{An-steps}
  \exp\left\{{\cal A}_{\Gamma^{(n)}_{J;K}}-{\cal A}_{\mathbb{R}}\right\} 
    = \prod_{m=1}^n\left(
  \frac{\alpha_-\bigl(\qm_{k_m};\Gamma^{(m-1)}_{j_1,\dots,j_{m-1};k_1,\dots,k_{m-1}}\bigr)}
 {\alpha_+\bigl(\qp_{j_m};\Gamma^{(m-1)}_{j_1,\dots,j_{m-1};k_1,\dots,k_{m-1}}\bigr)}
 \right)^2
 \frac{e_+^2(\qp_{j_m})e_-^2(\qm_{k_m})}{(\qp_{j_m}-\qm_{k_m})^2\gamma^2
 F'(\qp_{j_m})F'(\qm_{k_m})}.
 \end{equation}
It follows from \eqref{ar-ag} that
 \begin{equation}\label{ag2-a0}
 \alpha_\pm\bigl(\lambda;\Gamma^{(m-1)}_{j_1,\dots,j_{m-1};k_1,\dots,k_{m-1}}\bigr)
 =\alpha_\pm(\lambda)\prod_{a=1}^{m-1}
 \left(\frac{\lambda-\qm_{k_a}}{\lambda-\qp_{j_a}}\right).
 \end{equation}
It remains to substitute this into \eqref{An-steps} and to use the
equation
 \begin{equation}\label{Cauchy-det}
 \det_{j_1,\dots,j_n\atop{k_1,\dots,k_n}}\frac1{\qp_{j_a}-\qm_{k_b}}=
 \frac{\prod\limits_{a>b}^n
 (\qp_{j_a}-\qp_{j_b})(\qm_{k_b}-\qm_{k_a})}{\prod\limits_{a,b=1}^n
 (\qp_{j_a}-\qm_{k_b})}.
 \end{equation}
\qed

{\sl Proof  of Theorem~\ref{Main-thm-FD1}.} The asymptotic behavior
of the Fredholm determinant \eqref{logdet11} contains the
determinant $\det(I-A)$ with $A=A^-A^+$. Let for definiteness
$\np\le\nm$ (otherwise $\det(I-A)$ can be replaced by $\det(I-\tilde
A)$ with $\tilde A=A^+A^-$ due to \eqref{detA}). Then one has
 \begin{equation}\label{trans-detA1}
 \det_{~~\np}(I-A)=\sum_{n=0}^{\np}\sum_{j_n>\dots>j_1}^{\np}
 \det_{a,b=1,\dots,n}\left(-\sum_{k=1}^{\nm}A^-_{j_a,k}A^+_{k,j_b}\right)
 \end{equation}
Substituting here $A^\pm$ from \eqref{A-A+A} we obtain
 \begin{equation}\label{trans-detA2}
 \det_{~~\np}(I-A)=\sum_{n=0}^{\np}\sum_{j_n>\dots>j_1}^{\np}
 \sum_{k_1,\dots,k_n=1}^{\nm}\prod_{a=1}^n
 \frac{h^-_{k_a}h^+_{j_a}e_-^2(\qm_{k_a})e_+^2(\qp_{j_a})}
 {\qp_{j_a}-\qm_{k_a}}\det_{a,b=1,\dots,n}\left(\frac1{\qp_{j_a}-\qm_{k_b}}\right)
 \end{equation}
It remains to use the explicit expressions \eqref{A-A+A} for $h^\pm$
and to make the replacement
 \begin{equation}\label{repl-sum-det}
 \prod_{a=1}^n
 (\qp_{j_a}-\qm_{k_a})^{-1}\mapsto
 \frac1{n!}\det_{a,b=1,\dots,n}\left(\frac1{\qp_{j_a}-\qm_{k_b}}\right).
 \end{equation}
Such the replacement is possible, since we take the sum over all
$j_\ell$ and $k_\ell$. Then due to \eqref{JGR-fin1} we arrive at
 \begin{equation}\label{trans-detA3}
 \det_{~~\np}(I-A)=\sum_{n=0}^{\np}\sum_{j_n>\dots>j_1}^{\np}
 \sum_{k_n>\dots>k_1}^{\nm}\exp\left\{{\cal A}_{\Gamma^{(n)}_{J;K}}([g],[\nu])-{\cal
A}_{\mathbb{R}}([g],[\nu])\right\}.
 \end{equation}
Taking into account \eqref{logdet1} we obtain the statement of
Theorem.\qed

\section{Examples\label{Ex}}

In this section we consider applications of the results obtained.

\subsection{Temperature correlation function of impenetrable bosons\label{TCF-IB}}

The operator of number of particles $Q_x$ on an interval $[0,x]$
plays an important role in the theory of quantum one-dimensional
integrable systems \cite{IzeK84}--\cite{BogIKL93}. The expectation
value of $e^{\beta Q_x}$, where $\beta$ is a complex parameter, is a
generating function for some correlation functions of such models.
In the model of impenetrable bosons at finite temperature this
expectation value is given by the Fredholm determinant of the
operator $I+\gamma V$ \eqref{IGSK} with $g(\lambda)=0$ and
 \begin{equation}\label{gF}
 \gamma F(\lambda)=\frac{e^\beta-1}{e^{\frac{\lambda^2-h}T}+1},
 \end{equation}
where $h$ is the chemical potential and $T$ is the temperature. The
leading terms of the large $x$ asymptotic expansion of this
determinant were calculated in \cite{ItsIK90,ItsIK90a,BogIKL93}. The
formulas \eqref{logdet1}, \eqref{logdet0} give the complete
asymptotic expansion. Hereby
 \begin{equation}\label{alp-T}
 \nu(\lambda)=\frac{-1}{2\pi i}
 \log\left(\frac{e^{\frac{\lambda^2-h}T}+e^\beta}{e^{\frac{\lambda^2-h}T}+1}
 \right),
 \end{equation}
and the function $\alpha(\lambda)$ is given by \eqref{nu-1}. The
roots of the equation $1+\gamma F(\lambda)=0$ form two series
$\qpm_{j,i}$, $i=1,2$:
 \begin{equation}\label{roots1}
 \qp_{j,1}=\sqrt{h+\beta T+i\pi T(2j+1)},\qquad j=0,1,\dots\;,
 \end{equation}
and $\qp_{j,2}=- (\qp_{j,1})^*$, $\qm_{j,1}= (\qp_{j,1})^*$,
$\qm_{j,2}=-(\qm_{j,1})^*$. The equations  \eqref{logdet1},
\eqref{logdet0} are valid for arbitrary $N=\np=\nm$. For $N$ fixed
we neglect the roots $\qpm_{N+1}$, whose contribution is of order
$e^{ix(\qp_{N+1,j}-\qm_{1,k})}$. Therefore one can set
$a=\Im(\qp_{N+1,1}-\qm_{1,1})$ in the reminder $O(e^{-ax})$.

\subsection{Normalization factor in the $XXZ$ spin chain\label{norm}}

Our second example is related to the $XXZ$ spin-$1/2$ Heisenberg
chain in an external magnetic field. In the thermodynamic limit
correlation functions of this model contain a normalization factor
proportional to the Fredholm determinant ${\det}_{[-x/2,x/2]}(I+K)$
\cite{Kor82}, where the operator $K(t-t')$ acts on the interval
$[-x/2,x/2]$ and has the kernel
 \begin{equation}\label{K-kern}
 K(t-t')=\frac{\sin2\zeta}{2\pi\sinh(t-t'-i\zeta)\sinh(t-t'+i\zeta)}.
 \end{equation}
Here $0<\zeta<\pi$ and $\cos\zeta=\Delta$, where $\Delta$ is the
anisotropy parameter of the model. The length of the interval $x$
depends on the value of the external magnetic field. If the last one
goes to zero, then  $x\to\infty$.

The operator $I+K$ belongs to the class of truncated Wiener--Hopf
operators. They  can be reduced to the operators with the GSK by the
Fourier transform. Let $\chi_{[-\frac x2,\frac x2]}(t')$ be the
characteristic function of the interval $[-x/2,x/2]$. Then
 \begin{equation}\label{Rav-det}
 {\det}_{[-\frac x2,\frac x2]}(I+K)={\det}_{\mathbb{R}}(I+V),
 \end{equation}
where the operator $I+V$ acts on $\mathbb{R}$ and
 \begin{equation}\label{V-Four}
 V(\lambda,\mu)=\frac1{2\pi}\int\limits_{\mathbb{R}}e^{it\lambda}K(t-t')\chi_{[-x/2,x/2]}(t')
 e^{-it'\mu}\,dt\,dt'.
 \end{equation}
Calculating the integral in \eqref{V-Four} we find
 \begin{equation}\label{V-Four1}
 V(\lambda,\mu)=F(\lambda)\frac{\sin
 \frac{x}2(\lambda-\mu)}{\pi(\lambda-\mu)},
  \end{equation}
where
 \begin{equation}\label{F-kern}
 F(\lambda)=\hat K(\lambda)=\frac{\sinh[\lambda(\pi/2-\zeta)]}{\sinh[\lambda\pi/2]}.
 \end{equation}
Thus, up to the similarity transformation we obtain the GSK with
$g(\lambda)=0$ and $F(\lambda)$ given by \eqref{F-kern}. Hence, if
$x$ is large enough (what corresponds to a small magnetic field), we
can calculate the original determinant ${\det}_{[-x/2,x/2]}(I+K)$
asymptotically.

Just like in the previous example the formulas \eqref{logdet1},
\eqref{logdet0} give the complete asymptotic expansion. We have
 \begin{equation}\label{F-kern+1}
 \nu(\lambda)=\frac{-1}{2\pi i}
 \log\left(\frac{2\sinh[\lambda(\pi-\zeta)/2]\cosh[\lambda\zeta/2]}{\sinh[\lambda\pi/2]}
 \right).
 \end{equation}
The function $\alpha(\lambda)$ can be found explicitly in terms of
$\Gamma$-functions
 \begin{equation}\label{alp-expl}
 \alpha_-(\lambda)=\sqrt{2(\pi-\zeta)}\left(\frac{\pi}{\zeta}\right)^{-\frac{i\lambda\zeta}{2\pi}}
 \left(\frac{\pi}{\pi-\zeta}\right)^{-\frac{i\lambda(\pi-\zeta)}{2\pi}}
 \frac{\Gamma\left(1+\frac{i\lambda}{2}\right)}{\Gamma\left(\frac12+\frac{i\lambda\zeta}{2\pi}\right)
 \Gamma\left(1+\frac{i\lambda(\pi-\zeta)}{2\pi}\right)},
 \end{equation}
and $\alpha_+^{-1}(\lambda)=\alpha_-(-\lambda)$. Similarly to the
case considered above the roots of the equation $1+\gamma
F(\lambda)=0$ form two series  $\qpm_{j,i}$, $i=1,2$. If $\pi/\zeta$
is not a rational number, then
 \begin{equation}\label{roots2}
 \qp_{j,1}=\frac{2\pi i}{\pi-\zeta}(j+1),\qquad
 \qp_{j,2}=\frac{\pi i}{\zeta}(2j+1),\qquad j=0,1,\dots \;,
 \end{equation}
and $\qm_{j,i}= (\qp_{j,i})^*$. If $\pi/\zeta$ is  a rational
number, then one should omit in \eqref{roots2} such $\qpm_{j,i}$
that satisfy the condition $\sinh(\pi\qpm_{j,i}/2)=0$. The
asymptotic estimates \eqref{logdet1}, \eqref{logdet0} are valid for
arbitrary $N=\np=\nm$. Similarly to the case considered above one
can set $a=\min_{i,i'=1,2}\Im(\qp_{N+1,i}-\qm_{1,i'})$ in  the
reminder is $O(e^{-ax})$.

\section*{Acknowledgements}

I am grateful to N.~Kitanine, K.K.~Kozlowski, J.M.~Maillet and
V.~Terras for useful discussions. This work was supported in parts
by French-Russian network in Theoretical and Mathematical Physics
(GDRI-471 of CNRS and RFBR-CNRS-09-01-93106L-a), the Program of RAS
``Mathematical Methods of the Nonlinear Dynamics'',
RFBR-08-01-00501a, RFBR-09-01-12150ofi-m, NS-8265.2010.1.


\end{document}